\documentstyle[12pt]{article}
\begin{document}

\newtheorem{theorem}{Proposition}

\begin{center}
{\Large {\bf Kinematical symmetries of
3D incompressible flows}} \\
\vspace{2cm}
{\large {Hasan G\"{u}mral}\footnote{hasan@gursey.gov.tr}   \\  
\vspace{5mm}
Feza G\"{u}rsey Institute  \\
P.O. Box 6, 81220 \c{C}engelk\"oy-\.Istanbul, Turkey} \\ 
\vspace{5mm}
\today
\end{center}

\vspace{15mm}

\section*{Abstract}
The motion of an incompressible fluid in Lagrangian coordinates
involves infinitely many symmetries generated by the left Lie
algebra of group of volume preserving diffeomorphisms
of the three dimensional domain occupied by the fluid.
Utilizing a $1+3$-dimensional Hamiltonian setting an explicit
realization of this symmetry algebra is constructed recursively.
A dynamical connection is used to split the symmetries into
reparametrization of trajectories and one-parameter family of
volume preserving diffeomorphisms of fluid domain.
Algebraic structures of symmetries and Hamiltonian structures
of their generators are inherited from the same construction.
A comparison with the properties of 2D flows is included.

\newpage

\section{Introduction}

This paper presents a systematic approach, in the framework of
symplectic geometry, to the construction of infinitely many
symmetries of three-dimensional incompressible
hydrodynamic flows in the Lagrangian description.
The main construction presented in sections(\ref{symmtry})-(\ref{hss})
and summarized in section(\ref{summ}) relies on connecting the
automorphisms of a symplectic structure on the time-extended
space $I \times M$ to the symmetries of the time-dependent
motion on $M$.
The symmetries are consequences of the description of
motion and thus, are kinematical.
The corresponding invariants are related to the construction
of reduced phase space for the reduced Eulerian dynamical
equations.

\subsection{Kinematical symmetries and reduction}

We shall describe briefly those aspects of the problem of
orbits of coadjoint action that the content of this work
refers to. We shall do this by comparing the fluid motion
with the motion of a finite dimensional system of rigid body.

The configuration space of an incompressible fluid is the
group $Diff_{vol}(M)$ of volume preserving transformations
of the three dimensional
region $M$ in $R^3$ containing the fluid.
The motion is generated by the left action of the group on $M$
by evaluation and hence the velocity field is right invariant
(cf. section(\ref{mot})).
Thus, the generators of the right action which form the infinite
dimensional left Lie algebra of $Diff_{vol}(M)$ are infinitesimal
symmetries of the velocity field \cite{arnold}-\cite{mr}.
In the fluid mechanical context, these are known as the particle
relabelling symmetries \cite{mwe}-\cite{schutz}
also named as gauge transformations \cite{schutz},\cite{hen},
trivial displacements \cite{schutz} or
invisible symmetries \cite{shep}.

It was first shown in \cite{arnold} and described 
in a rigorous infinite dimensional context in \cite{mw}
that the dynamical formulation of motion of
an ideal fluid can be performed on exactly the same footing
as the system of a rigid body with one point fixed,
that is, the Euler top. The configuration space of this
finite dimensional system is the group $SO(3)$ of rotations.
In both cases one can obtain, via Lie-Poisson reduction, a proper
description of dynamics on the orbits of coadjoint action
of the group on the dual of its Lie algebra
\cite{mw},\cite{via}-\cite{mr}.

For the Euler top, the $SO(3)$ symmetry of the motion gives
a Casimir for the Hamiltonian description on the reduced
momentum phase space $so(3)^*$. The intersection of Casimir and energy
surfaces gives the geometric construction of the solution
space for the reduced dynamics.
The description of the reduced phase space for the Eulerian
equations of hydrodynamics is, however, rather difficult.
As suggested by the geometry of the reduced phase space
of finite dimensional examples like rigid bodies
\cite{via}-\cite{mr},\cite{shep}-\cite{wei} and
Volterra systems \cite{fer},
this possibly involves intersections of infinitely many
Casimirs of motion related to the particle relabelling
symmetries \cite{via},\cite{em},\cite{ebin}.

It has been the subject of many investigations to discover the
explicit and the most general form of the invariants 
of ideal fluid corresponding to the infinite dimensional
group of kinematical symmetries of motion
\cite{salmon}-\cite{hen},\cite{cal}-\cite{arsin}.
In spite of the fact that the particle relabelling
symmetry is a direct consequence of the description of motion itself,
most of the works on the structure of symmetries and
invariants have employed the techniques of analysing the defining
equations for them thereby obtaining an algebraically and
quantitatively incomplete picture of symmetries, invariants
and their connections.
\cite{schutz},\cite{hen},\cite{sagdeev},\cite{kuzmin}.
Moreover, although there is no essential difference in group
theoretical description of motion for two and three dimensional
flows, qualitatively different results for the corresponding
symmetries and the associated invariants were obtained
(cf. section(\ref{com}))
\cite{shep},\cite{khch}-\cite{arsin}.

In this work we shall present a geometric framework for a
systematic study of symmetries and invariants which will
provide, by addressing explicitly to Lagrangian and Eulerian
formulations, a better understanding of them in the description
of motion. We shall show that the resulting framework is
appropriate for the realization in three dimensions of 
some geometric properties of two-dimensional flows as well.

More background material and additional topics can be found in
\cite{via}-\cite{mr},\cite{mw}-\cite{wei},\cite{cm},
for Hamiltonian formalisms, symmetries and
reductions with applications to the motion of ideal fluids, in
\cite{em}-\cite{mr},\cite{ebin},\cite{arsin}-\cite{es}
for infinite dimensional diffeomorphism groups and algebras
as well as their functional analytic properties.

\subsection{Overview and content}

We shall consider the Lagrangian description of fluid motion
in the framework of a formal symplectic structure to obtain a realization
of the symmetry algebra.
We shall start from the observation that the Euler equations
themselves are the conditions for the vorticity field to be
an infinitesimal symmetry of the velocity field $v$.
We shall use this symmetry to construct
a symplectic structure on the time-extended domain $I \times M$
with $I \subset R$ and $M \subset R^3$.
This will define the Hamiltonian structure of the suspended
velocity field $\partial_t+v$.
Identifying a Hamiltonian vector field $\xi \partial_t+u$
on $I \times M$ as a new symmetry of
the velocity field, we shall be able to generate recursively
an infinite hierarchy of symmetries.

The vorticity is, by construction, an automorphism of the
Hamiltonian structure. Since any Hamiltonian vector field
is also an automorphism, we shall conclude that the resulting
generators do form a Lie algebra over $I \times M$.
The interpretation of the dynamical system associated with the
velocity field as a connection on
$I \times M$ will split the algebra of Hamiltonian symmetries
into gauge algebra of reparametrization symmetries
of trajectories of the velocity field, and the algebra of
one-parameter family of volume
preserving diffeomorphisms on $M$.

After a brief description of fluid motion in the next section
we shall recall from Ref.\cite{hg97} the symplectic structure of the
suspended velocity field on the time-extended space $I \times M$.
In section(\ref{symmtry}), we shall obtain recursively an infinite
hierarchy of time-dependent symmetries on $I \times M$ of the
velocity field and elobarate the properties of generators.
In section(\ref{split}) we shall split the symmetries on $I \times M$
into reparametrizations and volume preserving diffeomorphisms of $M$.
In section(\ref{hss}) we shall present the Hamiltonian structures
of symmetry generators and establish the isomorphism between the
left Lie algebra of $Diff_{vol}(M)$ and the Poisson bracket algebra
of time-dependent invariant functions on $M$.
In section(\ref{com}) we shall compare these results with the
properties of two-dimensional flow and summarize them in section(\ref{summ}).

\section{Three dimensional fluid motion}

\label{mot}

Let the open set $M_0 \subset R^{3}$ be the domain occupied initially
by an incompressible fluid and 
$x(t=0)=x_{0} \in M_0$ be the initial position, i.e., a Lagrangian label.
For a fixed initial position $x_{0}$,
the Eulerian coordinates $x(t) = g_{t}(x_{0})$ define 
a smooth curve in $R^{3}$
describing the evolution of fluid particles. 
For each time $t \in I$, the volume preserving embedding 
$g_{t}:M_0 \to g_t(M_0)=M \subset R^{3}$ describes a
configuration of fluid. A flow is then a curve $t \mapsto g_{t}$ 
in the space of all such transformations.
The time-dependent Eulerian (spatial) velocity field $v_{t}$
that generates $g_{t}$ is defined by
\begin{equation}
    {dx \over dt}={dg_{t}(x_{0}) \over dt}
            =(v_{t} \circ g_{t})(x_{0})=v(t,x)    \label{vel}
\end{equation}
where $v_{t} \circ g_{t}$ is the corresponding Lagrangian (material)
velocity field \cite{via},\cite{mr},\cite{djr}.
Since $g_{t}$ is volume preserving, $v_{t}(x)$ is a divergence-free vector
field over $M$ and Eq.(\ref{vel}) is a non-autonomous dynamical system
associated with it.
The system (\ref{vel}) can, equivalently, be represented as an
autonomous system defined by the suspended velocity field
\begin{equation}
  \partial_{t} + v(t,x)   \;,\;\;\;\;\; v={\bf v} \cdot \nabla \label{tvel}
\end{equation}
on the time-extended space $I \times M$.

The Lagrangian description of fluid motion is the description
by trajectories \cite{djr},\cite{sw}-\cite{vishik},
that is, by solutions
of non-autonomous ordinary differential equations (\ref{vel}).
Equivalence of (\ref{vel}) to the autonomous system associated
with (\ref{tvel}) means that the trajectories can be obtained
by describing streamlines at each time.
In Ref.\cite{hg97}, using the Eulerian dynamical equations,
we constructed a formal symplectic structure for (\ref{tvel})
on a time-extended domain $I \times M$. We shall now summarize
this construction.

\subsection{Symplectic structure}

A symplectic structure \cite{via},\cite{mr},\cite{am},\cite{gs}
on a manifold $N$ of even dimension $2n$ is defined by a closed,
non-degenerate two-form $\Omega$. It is exact if there exists a 
one-form $\theta$ such that $\Omega =- d \theta$.
Darboux's theorem quarantees the existence of 
local coordinates $(q^{i},p_{i})\; i=1,...,n$ 
in which $\Omega$ has the canonical form $dq^{i} \wedge dp_{i}$.
The $2n-$form $(-1)^{n} \Omega^{n}/n!$ is called the Liouville volume.
A vector field $V$ on $N$ is called Hamiltonian if there exists
a function $h$ on $N$ such that 
\begin{equation}
      i(V)(\Omega)=dh                 \label{seq}
\end{equation}
where $i(V)(\cdot)$ denotes the inner product with $X$.
The identity $i(V)(dh)=0$ which follows from (\ref{seq}) is 
the expression for conservation of $h$ under the flow of $V$.
With the correspondence (\ref{seq}) between functions and vector fields,
the Poisson bracket of functions on $N$ defined by
\begin{equation}
   \{ f,g \} = \Omega (V_{f},V_{g}) = \Omega^{-1}(df,dg)   \label{pobi}
\end{equation}
satisfies the conditions of bilinearity, skew-symmetry, the Jacobi identity
and the Leibniz rule.
This enables us to write the dynamical system associated with the
vector field $V_{h}$ in the form of Hamilton's equations
\begin{equation}
   {dx \over dt} = \{ x,h  \} =V_h(x)       \;.      \label{heq}
\end{equation}

\begin{theorem}             \label{symplec}
In the Eulerian description of motion of an incompressible fluid
let the dynamics of the velocity field ${\bf v}$ be governed by
\begin{equation}
  {\partial {\bf v} \over \partial t} +
  {\bf v} \cdot \nabla {\bf v}= {\bf F}   \label{geuler}
\end{equation}
and assume that the divergence-free field ${\bf B}$ and the function
$\varphi$ satisfy
\begin{equation}
    {\partial {\bf B} \over \partial t} - 
    \nabla \times ( {\bf v} \times {\bf B} ) =0 \;,\;\;\;
    {\partial \varphi \over \partial t} +
     {\bf v} \cdot \nabla \varphi =0     \label{beq}
\end{equation}
which are the frozen field equations. Then

(1) $\partial_{t}+v$ is a Hamiltonian vector field with the symplectic
two-form
\begin{equation}
  \Omega = - (\nabla \varphi + {\bf v} \times {\bf B}) 
          \cdot d{\bf x} \wedge dt + {\bf B} \cdot 
             (d{\bf x} \wedge d{\bf x})    \label{symp2}
\end{equation}
and the Hamiltonian function $\varphi$.

(2) The invariant Liouville volume density
$\rho_{\varphi} \equiv - {\bf B} \cdot \nabla \varphi$
is associated with the symplectic volume
\begin{equation}
  \mu \equiv \Omega \wedge \Omega =
      \rho_{\varphi} dt \wedge dx \wedge dy \wedge dz \;.  \label{vol}
\end{equation}

(3) If moreover ${\bf B} = \nabla \times {\bf A}$ for some vector 
potential ${\bf A}$, then $\Omega$ is exact
\begin{equation}
 \Omega = -d \theta \;,\;\;\; 
        - \theta =  \psi \, dt + {\bf A} \cdot d{\bf x}     \label{cone2}
\end{equation}
where $\psi$ is determined by the equation 
\begin{equation}
  {\partial {\bf A} \over \partial t} -
  {\bf v} \times (\nabla \times {\bf A}) = 
  \nabla (\varphi + \psi)     \;.            \label{aeq}
\end{equation}

(4) In Darboux coordinates,
$\Omega = dq \wedge dp + dt \wedge dh_{can}$ and the potential field
has the Clebsch representation \cite{mwe},\cite{sw},\cite{km}
\begin{equation}
  {\bf A}= \nabla s  + p \nabla q
\end{equation}
where $s$ is the generating function of the canonical transformation.
\end{theorem}

The Euler flow of an ideal isentropic fluid
is characterized by ${\bf F}= -\nabla(P/ \rho)$, where
$P$ and $\rho$ are pressure and density, respectively.
In this case, ${\bf A}$ is replaced by the velocity
field itself, ${\bf B}$ becomes the vorticity
${\bf w}=\nabla \times {\bf v}$ and Eqs.(\ref{aeq}) reduces
to the expression
\begin{equation}
    - \psi = \varphi + P / \rho +v^2 /2       \label{psio}
\end{equation}
for the scalar potential. The invariant volume density
$\rho_{\varphi} =- {\bf w} \cdot \nabla \varphi$ 
is also known as potential vorticity \cite{hen},\cite{sew}. In this case,
the frozen-field equation for ${\bf w}$ in (\ref{beq})
is equivalent to the Lie-Poisson equations for (\ref{geuler})
\cite{mg}-\cite{mor} on the dual
${\cal X}^*_{div}(M)$ of the Lie algebra of $Diff_{vol}(M)$
\cite{mwe}-\cite{mr}.

The main assumption in proposition(\ref{symplec}) is the existence
of a vectorial frozen-in field for the flow. This is in agreement
with the argument that the geometry of the velocity field
is determined by fields advected by its flow \cite{mof}.
The construction of proposition(\ref{symplec}) may be regarded
to be the analog of the determination of the symplectic structure
of a relativistic particle by electromagnetic field \cite{gs}.
In \cite{hg97} we presented a counter-example from dynamo theory
\cite{soward} to illustrate the necessity of frozen-in condition
in the construction
of the symplectic structure of suspended velocity field.
The physical framework of proposition(\ref{symplec})
includes the equations of non-relativistic
superconductivity \cite{feyn},\cite{hoku},\cite{khch},\cite{arsin},
the ideal magnetohydrodynamic equations
\cite{khch},\cite{arsin},\cite{kuzmin},\cite{zk},
barotropic fluids (pressure depends only on density)
\cite{khch},\cite{hmrw},\cite{arsin},
equations for perfectly conducting viscous
and finitely conducting inviscid fluids. 
This class of equations can be enlarged by the so-called
pseudo-advection technique \cite{vcs} where the frozen-field
is moving with (i.e. an invariant of) the modified velocity
${\bf v}+ \nabla \times {\bf a}$ for some vector field ${\bf a}$.

\section{Infinitesimal symmetries}

\label{symmtry}

This relatively long section is the first part of the main
geometric construction. Here, we shall sketch the flow
of ideas to be presented in this and the next two sections.

The frozen-in field is an infinitesimal symmetry of the velocity
field. Moreover, it is an infinitesimal automorphism of the
symplectic structure. Any Hamiltonian vector field also generates
an infinitesimal automorphism. Requiring this to be an infinitesimal
symmetry of the velocity field relates their Hamiltonian
functions. Under certain conditions an infinite hierarchy in
the subalgebra ${\cal X}_{ham}(I \times M)$ of
Hamiltonian vector fields in ${\cal X}(I \times M)$
can be realized as infinitesimal symmetries of $v$.

When the dynamical system (\ref{vel}) is viewed as a section
$\partial_t+v$ of the first jet bundle over $I \times M$
the induced connection $\Gamma = dt \otimes (\partial_t+v)$
on $I \times M \to I$ splits the infinitesimal symmetries into
the direct sum
\begin{equation}
  {\cal X}_{ham}(I \times M) =
   gau(I \times M) \oplus {\cal X}(M)      \label{alg}
\end{equation}
of the gauge algebra of reparametrizations of trajectories and
of diffeomorphisms of $M$. Requiring the latter to preserve
the volume on $M$ induced by the Liouville volume $\mu$
on $I \times M$ results in
the generators of one-parameter family of particle relabelling
symmetries on $M$. The vector fields in each component of this
decomposition are Hamiltonian.
The main results of sections(\ref{symmtry})-(\ref{hss}) will be
compared with the properties of two dimensional flows in
section(\ref{com}) and be summarized in section(\ref{summ}).

\subsection{A recursive construction of symmetries}

A time-dependent vector field $U = \xi \partial_{t}+u$ on $I \times M$
is an infinitesimal geometric symmetry of the Lagrangian motion on $M$
described by $v$ if the criterion
\begin{equation}
  [ \partial_{t}+v, \xi \partial_{t}+u ] = 
                (\xi_{,t}+v(\xi))(\partial_{t}+v)      \label{symm}
\end{equation}
is satisfied. These are the most general symmetries of the system
(\ref{vel}) of first order ordinary differential equations \cite{olver}.
It follows from the identity
\begin{equation}
   \nabla \times ({\bf v} \times {\bf B})=
   ({\bf B} \cdot \nabla) {\bf v}-({\bf v} \cdot \nabla) {\bf v}
   + (\nabla \cdot {\bf B}) {\bf v} - (\nabla \cdot {\bf v}) {\bf B}
\label{idb}      \end{equation}
and the divergence-free conditions on ${\bf B}$ and ${\bf v}$ that
the frozen field equation (\ref{beq}) for ${\bf B}$ is an
expression for it to be an infinitesimal symmetry of $v$ for
which the right hand side of eqs(\ref{symm}) vanish.
It also follows from eqs(\ref{symm}) that an infinitesimal symmetry
takes a Lagrangian conserved quantity to another one.
Thus, the Liouville volume
density $\rho_{\varphi}$ is a conserved function, and hence
$\rho_{\varphi}^{-1} {\bf B}$ is also an infinitesimal symmetry.
We observe that this initial symmetry leaves the symplectic
two-form (\ref{symp2}) invariant. More precisely, one checks that
the Lie derivative of $\Omega$ with respect to the normalized
field $\rho^{-1}_{\varphi}{\bf B} \cdot \nabla$ vanishes.

Our intention now is to find another infinitesimal invariance
of $\Omega$ which generates a symmetry of the velocity field as well.
One of the best candidates for this is a Hamiltonian vector field
because the symplectic two-form is invariant under the flows of 
Hamiltonian vector fields.
This can easily be seen from the identity
\begin{equation}
  {\cal L}_{U}=i(U) \circ d+d \circ i(U)   \label{cart}
\end{equation}
for the Lie derivative
together with Eq.(\ref{seq}) and the closure of $\Omega$.
If $U$ is such a vector field, called an infinitesimal
automorphism of $\Omega$, then it follows from the identity
\begin{equation}
{\cal L}_{[ U, \rho^{-1}_{\varphi}B ]}=
  {\cal L}_{U} \circ {\cal L}_{ \rho^{-1}_{\varphi}B } -
  {\cal L}_{ \rho^{-1}_{\varphi}B }  \circ {\cal L}_{U}    \label{iden}
\end{equation}
that $[ U, \rho^{-1}_{\varphi}B ]$ also leaves $\Omega$ invariant.
Replacing $ \rho^{-1}_{\varphi}B $ with $[ U, \rho^{-1}_{\varphi}B ]$
in the identity (\ref{iden}) we see that one can generate an
infinite dimensional algebra of Hamiltonian vector fields
(over simply connected domains $M$ of fluid) as
invariants of the symplectic two-form.

Thus, if we can find $U$ which
is also an infinitesimal symmetry of the velocity field,
this so-called algebra of symplectic
automorphisms of $\Omega$ will be carried over to 
the symmetry algebra of the velocity field $v$. The identity (\ref{iden})
will then become the Jacobi identity
\begin{eqnarray}
 [ [ U, \rho^{-1}_{\varphi}B ] , \partial_t+v]&=&
  [ U, [ \rho^{-1}_{\varphi}B, \partial_t+v]] -
  [ \rho^{-1}_{\varphi}B , [ U , \partial_t+v]]     \label{viden}  \\
  &=& - [ \rho^{-1}_{\varphi}B , [ U , \partial_t+v]]   \label{vide}
\end{eqnarray}
of the algebra of vector
fields and will enable us to obtain the generators recursively.
\begin{theorem} \label{symmetry}
The Hamiltonian vector field
\begin{equation}
   U= \rho^{-1}_{\varphi} [ - B(h) (\partial_{t} +v) + {dh \over dt} B 
    + \nabla \varphi \times \nabla h  \cdot \nabla  ]  \label{fluid}
\end{equation}
associated with the symplectic two-form $\Omega$ and
the arbitrary smooth function $h$ is an infinitesimal symmetry of $v$ if 
$dh/dt=f(\varphi)$ for some function $f$, where $\varphi$ is
the Hamiltonian function for the suspension $\partial_t+v$.
In this case, the brackets
\begin{equation}
        [ U,..., [ U, [ U,\rho^{-1}_{\varphi} B ] ]...]    \label{infsym}
\end{equation}
generate an infinite hierarchy of time-dependent infinitesimal
Hamiltonian symmetries of the velocity field $v$.
\end{theorem}
{\bf Proof:}
For a given function $h$ the vector field (\ref{fluid}) is
uniquely determined by the Hamilton's equations (\ref{seq}).
For the symmetry condition on $U$ we compute the Lie bracket
\begin{equation}
  [ \partial_{t}+v, U ] = 
  - \rho^{-1}_{\varphi} B({dh \over dt}) (\partial_{t}+v)
    +  {d^2h \over dt^2} \rho^{-1}_{\varphi} B
   + \rho^{-1}_{\varphi} \nabla \varphi \times
        \nabla {dh \over dt}  \cdot \nabla
\end{equation}
and comparing with eq(\ref{symm}) we conclude that the sum
\begin{equation}
    {d^2h \over dt^2} \rho^{-1}_{\varphi} B
   + \rho^{-1}_{\varphi} \nabla \varphi \times
        \nabla {dh \over dt}  \cdot \nabla         \label{hbb}
\end{equation}
must vanish. Evaluation of this on $\varphi$ implies the invariance
of $dh / dt$ which must functionally depend on $\varphi$ for the
vanishing of the second term.

The recursive construction (\ref{infsym}) of symmetries
is obtained from the Jacobi identity (\ref{viden}) by
repeatedly performing the replacement
$ \rho^{-1}_{\varphi}B \mapsto  [ U, \rho^{-1}_{\varphi}B ]$.

The Hamiltonian property of infinitesimal symmetries
(\ref{infsym}) follows from the observation that
$U$ is chosen to be Hamiltonian and that $\rho^{-1}_{\varphi}B$
is a Hamiltonian vector field of $\Omega$
with the function $t$.
$\bullet$

In section(\ref{hss}), we shall present the Hamiltonian structures
of symmetries in more detail. To this end, we want to investigate
the structure of infinitesimal symmetries on $I \times M$
and their relation with the generators of the group $Diff_{vol}(M)$
of volume preserving diffeomorphisms of $M$.
We begin with the following main computational results which will be
the most relevant in foregoing discussions.
\begin{theorem}          \label{wvk}
For $h, \varphi$ being the Hamiltonian functions of $U, \partial_t+v$,
respectively, and $dh/dt=f(\varphi)$ for some arbitrary function $f$,
consider the time-dependent vector fields
\begin{equation}
 U_0 \equiv \rho^{-1}_{\varphi} B \;,\;\;\;
 W^v_1 \equiv \rho^{-1}_{\varphi}  \nabla \varphi
   \times \nabla h  \cdot \nabla  
\end{equation}
on $M \subset R^3$ involving the expression (\ref{fluid}) of the
new infinitesimal symmetry $U$.
Then, the time-dependent functions and vector fields
\begin{equation}
  h_k \equiv  -(W^v_1)^{k-2}(U_0(h)) \;,\;\;\;
  W^v_k  \equiv  ({\cal L}_{W^v_1})^{k-1}(U_0) \;,\;\; k \geq 2
        \label{hwk}
\end{equation}
are related through
\begin{equation}
  W^v_k = \rho^{-1}_{\varphi}  \nabla \varphi
   \times \nabla h_k  \cdot \nabla
   \;,\;\;\; k=2,3,4,...           \label{vert}
\end{equation}
and these satisfy the Lie bracket relations
\begin{equation}
  [ W^v_k, W^v_l ] =  \rho^{-1}_{\varphi} \nabla \varphi \times
  \nabla h_{lk} \cdot \nabla  \equiv W^v_{kl}   \label{wkl}
\end{equation}
where the functions
\begin{equation}
    h_{lk} \equiv \rho^{-1}_{\varphi}  \nabla \varphi \cdot
          \nabla h_l \times \nabla h_k
\end{equation}
and hence $W^v_{kl}$ are antisymmetric in their indices.
$U_0,W^v_k$ and $W^v_{kl}$ are time-dependent vector fields
on $M$ commuting with the suspension $\partial_t+v$
\begin{equation}
      [\partial_t+v,U_0]=[\partial_t+v,W^v_k]=
      [\partial_t+v,W^v_{kl}]=0   \;.          \label{bbb}
\end{equation}
\end{theorem}
{\bf Proof:}
We shall obtain the equation (\ref{vert}) by induction.
From the identity (\ref{idb}) we get
\begin{equation}
  [W_1^v,U_0]= \nabla \times ({\bf U}_0 \times {\bf W}^v_1 ) \cdot \nabla
   + \rho_{\varphi} \nabla \rho^{-1}_{\varphi} \times
   ( {\bf W}^v_{1} \times {\bf U}_0)  \cdot \nabla    \label{bra1}
\end{equation}
where we used the three-dimensional vector identity
\begin{equation}
 {\bf a} \times ({\bf b} \times {\bf c}) =
 ({\bf c} \cdot {\bf a}) {\bf b}
 - ( {\bf b} \cdot {\bf a}) {\bf c}             \label{abc}
\end{equation}
for the divergence terms in (\ref{idb}). Using (\ref{abc})
again, we have
\begin{equation}
  {\bf W}^v_1 \times {\bf U}_0 = - \rho^{-1}_{\varphi}
   ( \nabla h + U_0(h) \varphi)    \label{cross}
\end{equation}
and this gives for the bracket in (\ref{bra1})
\begin{equation}
  [W_1^v,U_0]= - \rho^{-1}_{\varphi}
    \nabla \varphi \times \nabla U_0(h)  \cdot \nabla
    \equiv  W_2^v \;.
\label{uv2}        \end{equation}
Assuming (\ref{vert}) is true, we repeat the same computation for
$k \geq 2$
\begin{eqnarray}
  {\bf W}^v_1 \times {\bf W}^v_k &=&
   {\bf W}^v_1 \times ( \rho^{-1}_{\varphi}
     \nabla \varphi  \times  \nabla h_k )       \\
      &=& \rho^{-1}_{\varphi}  W^v_1(h_k) \nabla \varphi
         \; = \; \rho^{-1}_{\varphi}  h_{k+1} \nabla \varphi
       \label{crossk}
\end{eqnarray}
and eq(\ref{bra1}) with $U_0$ replaced by $W^v_k$ implies
\begin{equation}
  [ W^v_1, W^v_k ] =  \rho^{-1}_{\varphi}
  \nabla \varphi \times \nabla h_{k+1} \cdot \nabla
  = W^v_{k+1} \;.
\label{uvk1}       \end{equation}
The bracket relation (\ref{wkl}) also follows from eq(\ref{bra1})
and
\begin{eqnarray}
  {\bf W}^v_l \times {\bf W}^v_k &=&
   {\bf W}^v_l \times ( \rho^{-1}_{\varphi}
     \nabla \varphi  \times  \nabla h_k )       \\
      &=& \rho^{-1}_{\varphi}
      (\rho^{-1}_{\varphi} \nabla \varphi \times \nabla h_l
      \cdot \nabla h_k) \nabla \varphi
         \; = \; - \rho^{-1}_{\varphi}  h_{lk} \nabla \varphi  \;.
       \label{crossk}
\end{eqnarray}
The commutativity of $W^v_k$ with $\partial_t+v$ follows from the
identity (\ref{viden}) once we have $[W^v_1, \partial_t+v]=0$
which amounts to the vanishing of the second term in eq(\ref{hbb}).
 $\bullet$

With the help of these results
the recursive construction of proposition(\ref{symmetry}) gives
\begin{eqnarray}
 U_0&=& \rho^{-1}_{\varphi} B \\
 U_1&=&U\;=\; h_2 (\partial_{t} +v) + f U_0 + W^v_1  \\
 U_2&=& - U_0(h_2) (\partial_{t} +v) + f' U_0 + W^v_2   \\
 U_3&=& \{ f''h_2 - f (U_0)^2(h_2) - U_0(h_3)  \}
             (\partial_{t} +v)  \nonumber   \\
    & & + \{ (f')^2-ff'' \}  U_0 
        + f {\cal L}_{U_0}(W^v_2) + f' W^v_2 + W^v_3  \\
    & &  \ldots       \nonumber
\end{eqnarray}
for the first few members of the hierarchy (\ref{infsym}).
Here, prime denotes the derivative of $f$ with respect to $\varphi$.
It can be seen by inspection that
the infinitesimal symmetries (\ref{infsym}) are of the form
\begin{equation}
   U_{k} = \xi_k (\partial_t +v) + U^v_k
\end{equation}
where for notational consistency with the next section we denote
\begin{equation}
 U^v_k = a_k U_0 +
   \sum_{i+j=k} b_i ({\cal L}_{U_0})^j(W^v_i)
   + \sum_{i=2}^{k-1} c_i W^v_i
   + W^v_k
   \label{uk}  \end{equation}
for the time-dependent vector fields on $M$
which commute with $\partial_t+v$. The coefficients
$a_k,b_i$ and $c_i$ are conserved functions of the velocity
field whereas $\xi_k$ are not.
The repeated Lie derivatives of $W^v_k$ with respect to $U_0$ can be
computed to be
\begin{equation}
 ({\cal L}_{U_0})^k(W^v_l) = \rho^{-1}_{\varphi}  \nabla \varphi
   \times \nabla (U_0)^k(h_l)  \cdot \nabla     
\end{equation}
with $h_1=h$.
The Lie bracket relations (\ref{wkl}) and (\ref{bbb}) imply
that the formal procedure in proposition(\ref{symmetry}) can
be continued indefinitely by replacing $U=U_1$ with
$U_k,U_{kl},...$ and so on.

We shall show below that
the condition of preserving the volume on $M$ requires
$\xi_k$ to be conserved. This, in turn, implies the
conservation of $h$, that is, $f(\varphi)=0$ which reduces
the part $U^v_k$ of $U_k$ to the single term $W^v_k$.

\section{Splitting of symmetries}          \label{split}

We shall consider a framework which will relate the symmetries
on $I \times M$ obtained from the automorphisms of the symplectic
structure to the kinematical symmetries in $Diff_{vol}(M)$
of the time-dependent fluid motion on $M$.
We shall split the space of symmetries into
those preserving the trajectories of the velocity field $v$
and the time-dependent diffeomorphisms of $M$.
Restricting the generators $U^v_k$ of the latter by volume preserving
condition we shall obtain time-dependent divergence-free
vector fields $W^v_k$ on $M$ which are left invariant under the
adjoint action of $Diff_{vol}(M)$. The basic object to be employed
is a connection defined by the velocity field itself.

A time-dependent first order ordinary differential equation
on $M$ defines a subvariety in the first jet bundle over $I \times M$.
This coordinate free definition of the
dynamical system (\ref{vel}) as a section of the jet bundle
induces the connection
\begin{equation}
  \Gamma = dt \otimes (\partial_t +v)        \label{cnnc}
\end{equation}
on $I \times M$ defined by the velocity field $v$
\cite{olver},\cite{saun},\cite{cm}.
We find this interpretation to be appropriate for
the Lagrangian description of hydrodynamic motions
because the velocity field is defined only implicitly as solution of
some nonlinear Eulerian equations.

$\Gamma$ decomposes the tangent bundle $T(I \times M)$
into horizontal and vertical bundles
\begin{equation}
  T(I \times M)=  H(I \times M) \oplus V(I \times M)
\end{equation}
by its action on sections $U=\xi \partial_t +u \in {\cal X}(I \times M)$ of
$T(I \times M)$ as
\begin{eqnarray}
  U^h&=& \Gamma(U)\; =\; \xi (\partial_t+v) \; \in
                     {\cal X}^H(I \times M) \label{uh} \\
  U^v&=& U- \Gamma(U) \; = \;  u - \xi v \;  \in
                    {\cal X}^V(I \times M)  \label{decom}
\end{eqnarray}
where the superscripts $h$ and $v$ stand for the 
horizontal and vertical components, respectively.

The horizontal components $U^h \in {\cal X}^H(I \times M)$ generate
the reparametrizations $\tau(t,x)$ of the integral curves of the
velocity field $v$ as solution of the equation $d \tau = \xi(t,x) dt$.
These transformations which leave each trajectory invariant
are gauge transformations of the fluid motion.

The vertical components $U^v \in {\cal X}^V(I \times M)$ act on $M$ and
generate one-parameter family of diffeomorphisms which are
equivalent to the symmetries generated by (\ref{infsym}).
To see this, we have, for components of an
infinitesimal symmetry on $I \times M$ satisfying the
criterion (\ref{symm}) 
\begin{equation}
  [ \partial_{t}+v, U^h ] = (\xi_{,t}+v(\xi))( \partial_{t}+v) \;,\;\;\;
  [ \partial_{t}+v, U^v ] =0            \label{chsy}
\end{equation}
and one can check that the second equation
follows from Eq.(\ref{symm}) by direct computation. 
This form of the symmetry condition is appropriate for
the interpretation of the velocity field as one-jet field
defining a connection. In symmetry analysis of differential
equations $U^v$ is called the evolutionary representative
or characteristic form of $U$ \cite{olver}.

\begin{theorem}     \label{divf}
$U^v_k$'s are divergence-free with respect to the time-dependent
volume $i(\partial_t)(\mu)= \rho_{\varphi} dx \wedge dy \wedge dz$
on $M$ if and only if $h$ is conserved under the flow of $\partial_t+v$.
\end{theorem}
{\bf Proof:}
Since $U_k$'s are automorphisms of $\Omega$ they are divergence-free
with respect to the symplectic volume $\mu$.
If $U^h_k = \xi_k (\partial_t+v)$ is divergence-free,
so will be $U^v_k$. We compute
\begin{equation}
  {\cal L}_{\xi (\partial_t+v)}(\mu) = d \xi \wedge i(\partial_t+v)(\mu)
        + \xi div_{\mu}(\partial_t+v) \mu  = (\xi_{,t} +v(\xi)) \mu
\end{equation}
where we used $div_{\mu}(\partial_t+v)=0$ or equivalently, the
conservation of the Liouville density $\rho_{\varphi}$
and $\nabla \cdot {\bf v}=0$.
Thus, $U^h_k$ is divergence-free if $\xi_k$ is conserved.
Expressions for $\xi_k$ involve derivatives of $f$ with respect
to $\varphi$ which is conserved, and evaluation of various powers of
$U_0$ and $W^v_i$ for $i < k$ on the function $h$.
From the commutativity of $\partial_t+v$ with $U_0$ and $W^v_k$
we conclude that $d \xi /dt =0$ whenever
$dh/dt=f(\varphi)$ is conserved for $U_0$, that is, $U_0(\varphi)=0$.
But, the definition of the Liouville density
$\rho_{\varphi}=-B(\varphi) \neq 0$
implies $U_0(\varphi)=-1$. Hence, we must have $dh/dt=0$.
$\bullet$

Thus, the volume preserving condition on vertical components
led us to consider the subclass
\begin{equation}
   U_{k} = \xi_k (\partial_t +v) + W^v_k  \;,\;\;\; k=0,1,2,...
\label{uk}  \end{equation}
of the infinitesimal symmetries (\ref{infsym}) on $I \times M$
with $\xi_k$'s being conserved functions of $v$. We compute
\begin{equation}
  \xi_0=0 ,\; \xi_1 = -U_0(h) ,\; \xi_2 = - (U_0)^2(h) ,\;
\xi_3 = W^v_1((U_0)^2(h))= U_0(h_3),\;               \label{xii}
\end{equation}
to list a few of these conserved functions of the velocity field.
The apparent relation between $\xi_k, h_k$ and $U_0$ in eqs(\ref{xii})
is indeed generic. Since $U_0$ is Hamiltonian with the function $t$,
we have
\begin{equation}
  \xi_k = U_k(t) = i(U_k)i(U_0)(\Omega)
                   = -i(U_0)i(U_k)(\Omega) = -U_0(h_k) \;.
\end{equation}
The vertical components $W^v_k$ of the restricted hierarchy
(\ref{uk}) of infinitesimal symmetries are precisely the
left invariant generators of one parameter family of volume
preserving diffeomorphisms on $M$, that is, the particle relabelling
symmetries.
\begin{theorem}
The hierarchy (\ref{infsym}) of infinitesimal symmetries
on $I \times M$ induces the time-dependent generators
\begin{equation}
    [W^v_1,...[ W_1^v ,[ W_1^v,U_0 ]]...]    \label{infver}
\end{equation}
of volume preserving diffeomorphisms of $M$ that are elements
of the left Lie algebra of $Diff_{vol}(M)$.
In terms of $h, \varphi$ and $B$, the hierarchy (\ref{infver})
anchored to $U_0$ and $W^v_1$ consists of vector fields $W^v_k$
with expression and properties given in proposition(\ref{wvk}).
\end{theorem}
{\bf Proof:}
We shall show that the vector fields (\ref{infver}) are closed
under Lie bracket, divergence-free and left invariant.
The curvature $R_{\Gamma} : {\cal X}(I \times M) \times
{\cal X}(I \times M) \to {\cal X}(I \times M)$ of a connection
$\Gamma$ is defined by its values
\begin{equation}
 R_{\Gamma}(U,W)= [ U^h,W^h ]^v
       = (1- \Gamma)([ \Gamma(U), \Gamma(W) ])
\end{equation}
on arbitrary vector fields $U,W$. A straightforward computation
using eqs(\ref{uh}) and (\ref{decom}) for vector fields $U,W$ on
$I \times M$ shows that the curvature of $\Gamma$ vanishes. 
On the other hand, the Nijenhuis tensor 
$[ \Gamma, \Gamma ]_{FN} = N_{\Gamma}$ of $\Gamma$ computed via the 
Fr\"olicher-Nijenhuis bracket, is related to the curvature as
$R_{\Gamma}=N_{\Gamma} /2$ \cite{saun}. Thus, the
horizontal and vertical bundles are integrable in the sense
that their spaces of sections are closed under Lie bracket
of vector fields.
In symmetry analysis this result can be expressed by the statement
that the bracket of characteristic vector fields is the same as
the characteristic form of the bracket \cite{olver}.

The divergence-free condition for $W^v_k$ is discussed in
proposition(\ref{divf}).
Since $U_0$ is divergence-free with respect to the symplectic
volume, the result also follows from the identity
\begin{equation}
 div_{\mu}([U,W])=U(div_{\mu}(W)) - W(div_{\mu}(U))
\end{equation}
because $W^v_1$ is $\rho_{\varphi}$-divergence-free.

The left invariance of $W^v_k$ is the same as the symmetry
condition because it is the infinitesimal version of the
invariance of $W^v_k$'s under the pullback
$g^*_t(W^v_k) \equiv Tg_t^{-1} \circ W^v_k \circ g_t$
by the left action generated by the velocity field
and follows from the Lie derivative formula \cite{gs},\cite{mr}.
Hence, $W^v_k$'s belong to the left Lie algebra of $Diff_{vol}(M)$
and they generate the right action on $M$ which is the particle
relabelling symmetry.
 $\bullet$

\section{Hamiltonian structures of symmetries}       \label{hss}

We shall establish an isomorphism between the Lie bracket
algebra of vector fields $W^v_k$ in the left Lie algebra
of $Diff_{vol}(M)$ and the Poisson bracket algebra of
time-dependent left invariant functions on $M$. The Poisson bracket is
induced from and is compatible with the one on $I \times M$
associated with the symplectic structure. The set of functions
are Lagrangian invariants of fluid motion all of which are
in the form of potential vorticity.

\begin{theorem} The vector fields $U_k,W^v_k$ and $U^h_k$ are
Hamiltonian with the Poisson brackets
\begin{equation}
  \{ f,g \}_{\Omega} = 
  U_0(f){dg \over dt}- U_0(g){df \over dt} +
   \rho^{-1}_{\varphi}
  \nabla \varphi \cdot \nabla f \times \nabla g   \label{pobi}
\end{equation}
defined by the symplectic two-form (\ref{symp2}) on $I \times M$,
\begin{equation}
  \{ f,g \}_{\varphi} = \rho^{-1}_{\varphi} \nabla \varphi
  \cdot \nabla f \times \nabla g             \label{brm}
\end{equation}
characterized by the Hamiltonian function $\varphi$
of the suspension (\ref{tvel}), and
\begin{equation}
       \{ f,g \}_{-} = \{ f,g \}_{\Omega}- \{ f,g \}_{\varphi} =
  U_0(f){dg \over dt}-U_0(g){df \over dt}  \;,    \label{hgau}
\end{equation}
respectively. For each $k$ the Hamiltonian function
$h_k$ is common to all three.
The vector fields
\begin{equation}
  U_{kl}= [U_k,U_l] \;,\;\;\; W^v_{kl}= [W^v_k,W^v_l]
\end{equation}
are also Hamiltonian with the brackets (\ref{pobi}) and (\ref{brm}),
respectively. Their common Hamiltonian functions are given by
\begin{equation}
  h_{lk}= \{ h_l,h_k \}_{\Omega} = \{ h_l,h_k \}_{\varphi}
\end{equation}
and with respect to the bracket (\ref{hgau})
the functions $h_k$ form an involutive set or equivalently,
the gauge algebra of reparametrization symmetries is commutative.
\end{theorem}
{\bf Proof:}
The anti-symmetry of the brackets (\ref{pobi})-(\ref{hgau})
are obvious. The Jacobi identity
\begin{equation}
  \{ \{ f,g \} ,h \} + \{ \{ g,h \} ,f \} + \{ \{ h,f \} ,g \} =0
\end{equation}
for (\ref{pobi}) is equivalent to the closure of $\Omega$
\cite{mr},\cite{olver},\cite{via}.
For a bare hand proof of Jacobi identity for (\ref{hgau})
we compute, using $[d/dt,U_0]=[\partial_t+v,U_0]=0$
\begin{eqnarray}
  \{ \{ f,g \} ,h \} &=&
 - U_0(h)U_0({df \over dt}){dg \over dt}
 + U_0(h){df \over dt}U_0({dg \over dt} )       \nonumber  \\
 & & + {dh \over dt}U_0(f)U_0({dg \over dt} )
 -{dh \over dt}U_0({df \over dt} )U_0(g)
 - U_0(h)U_0(f){dg \over dt}             \nonumber      \\
 & & +U_0(h){d^2f \over dt^2}U_0(g)
 + {dh \over dt}U^2_0(f){dg \over dt}
 - {dh \over dt}{df \over dt}U^2_0(g)
\end{eqnarray}
and check that the cyclic permutations in the arbitrary functions
$f,g,h$ sum up to zero.

$U_0,U_1$ and hence $U_2= [ U_1,U_0 ] $ are Hamiltonian
vector fields by construction.
Assuming $U_k$ is Hamiltonian with $h_k$, we use the identity
\begin{equation}
  i([U,W]) = {\cal L}_U \circ i(W) - i(W) \circ {\cal L}_U    \label{idil}
\end{equation}
and eqs(\ref{hwk}) to obtain 
\begin{eqnarray}
  i(U_{k+1})(\Omega) &=& i([U_1,U_k])(\Omega)
 \; = \; {\cal L}_{U_1} i(U_k)(\Omega)    \nonumber    \\
 &= &  d U_1(h_k) \; = \; d W^v_1(h_k) \; = \; d h_{k+1}
\end{eqnarray}
which is the Hamilton's equation for the vector field $U_{k+1}$.

The Hamiltonian form of the vector fields $W^v_k$ are manifested
in eq(\ref{vert}). The Lie algebra isomorphism
\begin{equation}
   [ U_f, U_g ] = U_{ \{ f,g \} }             \label{isom}
\end{equation}
with the Poisson brackets (\ref{pobi}) and (\ref{brm}) implies
the Hamiltonicity of $U_{kl}$ and $W^v_{kl}$, respectively.
Since $h_k$'s are conserved functions of $v$, it follows from
(\ref{pobi}) and (\ref{brm}) that
$U_{kl}$ and $W^v_{kl}$ have common Hamiltonian functions
and $h_k$'s are involutive with respect to (\ref{hgau}).
It follows from the involutivity of $h_k$'s and the isomorphism
(\ref{isom}) for the bracket (\ref{hgau}) that the reparametrization
of trajectories form a commutative subgroup of the symmetries
generated by $U_k$'s. $\bullet$

The functions $h_k$ are conserved under the flow of the velocity field
and they are in the form of potential vorticity
\cite{hen},\cite{sew}. Thus, the bracket
(\ref{brm}) may be interpreted as to define the Poisson bracket algebra
of generalized potential vorticities on the flow space $M$ and
by (\ref{isom}) this algebra is isomorphic to the Lie bracket
algebra of vector fields $W^v_k$.

\section{Comparison with 2D incompressible flows}

\label{com}

The results presented so far can also be obtained for
two-dimensional unsteady flow of incompressible fluids
using a slightly different technique. In general case of
even dimensional flows,
one must start with a necessarily degenerate Poisson structure
on the odd dimensional manifold $I \times M$ and proceed similarly
in the framework of contravariant geometry \cite{wei}.
In fact, the present treatment with symplectic structure
is just a particular case of a general construction with
contravariant techniques.
Our aim in this section is to show that the time-extended
symplectic structure for three-dimensional flow provides
it with some well-known properties of two-dimensional flow.

In two dimensions, $M$ attains a symplectic structure defined
by its volume two-form and the problem acquires simplicity because
the algebra ${\cal X}_{div}(M)$ can be identified, via the
isomorphism (\ref{isom}) defined by the Hamiltonian structure,
with the canonical Poisson bracket algebra of nonconstant
functions on $M$ \cite{mwe}. More precisely,
a time-dependent, divergence-free velocity field $v$ on a two-dimensional
domain $M$ with coordinates $(x,y)$, and its curl vector field $w$
perpendicular to $M$ can be expressed by means of
a function $\psi = \psi (t,x,y)$ as
\begin{equation}
  v = { \partial \psi \over \partial x}
      { \partial \over \partial y}
     - { \partial \psi \over \partial y}
      { \partial  \over \partial x}   \;,\;\;\;
      w = \phi { \partial \over \partial z} \;,\;\;\;
      \phi \equiv \nabla^2 \psi          \label{2dv}
\end{equation}
and they satisfy the frozen-field equation for $w$
\begin{equation}
    { \partial \phi \over \partial t}
         + \{ \phi , \psi \}_{can}  =0          \label{ffw}
\end{equation}
where $\{ \;,\; \}_{can}$ is the canonical bracket on two-dimensional
domain $M$ defined by the volume two-form.
Eq(\ref{ffw}) is equivalent to the Euler equations in two dimensions.
We observe that the formal restriction of the symplectic two-form
(\ref{symp2}) to the vector fields (\ref{2dv}) manifests its interplay
with the two dimensional Eulerian dynamical equations. Namely, we find that
the degenerate two-form
\begin{equation}
   \Omega = - (d \varphi + \phi d \psi ) \wedge dt
             + \phi dx \wedge dy
\end{equation}
is closed whenever eq(\ref{ffw}) holds. Moreover, the suspended
velocity field $\partial_t+v$ in three dimensions
is Hamiltonian provided the Hamiltonian function $\varphi$
satisfies the same equation. Thus, we can identify $\phi$
as the Hamiltonian function.

Using the Lie algebra isomorphism (\ref{isom}) for the symplectic
structure in two dimensions,  
the $+$Lie-Poisson structure on ${\cal X}^*_{div}(M)$
can be written in vorticity form
\begin{equation}
  \{ F,G \} ( \phi ) =  \int_M \phi 
   \{   {\delta F \over \delta \phi },
       {\delta G \over \delta  \phi }  \}_{can} \; dx \, dy \label{pb2}
\end{equation}
which involves the canonical Poisson bracket on $M$. The functionals
\begin{equation}
  C ( \phi ) =  \int_M  \Phi(\phi) \; dx \, dy        \label{cas}
\end{equation}
are Casimirs of (\ref{pb2}) for arbitrary function $\Phi$ of
the Eulerian vorticity variable and are
always conserved for the Euler equations.
For three dimensional flows, on the other hand,
although the Casimirs \cite{ahmr},\cite{ah}
\begin{equation}
  C ( q, \rho ) =  \int_M  \Phi( q, \rho )  \; dx \, dy \, dz
\end{equation}
where $q=w(\rho)$ is the potential vorticity, also
involve an arbitrary function $\Phi$, contrary to $C(\phi)$,
$C(q,\rho)$ contain Lagrangian information \cite{shep}.
Moreover, the existence of infinitely many generalized enstropy
type invariants for even dimensional flows has been established
whereas for odd dimensional case, only the existence of a single
Casimir expressible in Eulerian variables,
namely, a generalized helicity is ensured
\cite{khch}-\cite{arsin}.

In this work, we showed that
the representation of area preserving diffeomorphisms
in two dimensions as Hamiltonian vector fields
can be realized, with appropriate modifications of geometric tools,
for the three dimensional flows as well.
Analogous to the bracket (\ref{pb2}) for two dimensional
flow, the Hamiltonian structure on $M$ of the vector fields
$W^v_k$ enables us to write the vorticity bracket in three dimensions
in terms of the Poisson bracket (\ref{brm}) on $M$ of
generalized potential vorticities.

\section{Summary, conclusions and prospectives}       \label{summ}

We presented a geometric approach to the explicit
construction of kinematical symmetries starting from an
initial one and utilizing a symplectic structure both of which
are implicit in the Eulerian equations constraining the velocity field.
We may conclude that,
as the simplicity and the Lie-Poisson structure
of Eulerian description derives from the particle
relabelling symmetry \cite{via}-\cite{salmon}, the Eulerian description,
in its turn, enables us to construct the Lie algebra of these symmetries.
We summarize
the main results on symmetries of three dimensional flow with
velocity field $v$ of an incompressible fluid admitting a frozen-in
field $B$ and a Lagrangian conserved quantity $\varphi$.
\begin{itemize}
\item
$\partial_{t}+v$ is a Hamiltonian vector field with the symplectic
two-form
$$  \Omega = - (\nabla \varphi + {\bf v} \times {\bf B}) 
          \cdot d{\bf x} \wedge dt + {\bf B} \cdot 
             (d{\bf x} \wedge d{\bf x})  $$
and the Hamiltonian function $\varphi$.

\item
The vector fields $U_k=({\cal L}_{U_1})^k(U_0)$ where
$U_0= \rho_{\varphi}^{-1} B$ and
$$  U_1= -U_0(h) (\partial_{t} +v) + f U_0 + W^v_1  \;,\;\;\;
 W^v_1 \equiv \rho^{-1}_{\varphi}  \nabla \varphi
   \times \nabla h  \cdot \nabla    $$
is the Hamiltonian vector field for 
$h$ satisfying $dh/dt=f(\varphi)$ for some function $f$,
are infinitesimal symmetries of $v$.

\item
The connection $\Gamma = dt \otimes (\partial_t+v)$
splits the vector fields $U_k=\xi_k \partial_t + u_k$ into generators
$U^h_k= \xi_k (\partial_t+v)$ of gauge transformations and
of diffeomorphisms $U^v_k =u_k - \xi_k v$ on $M$.

\item
If $h$ is a conserved function of $v$, so are $\xi_k= - U_0(h_k)$
where $h_k \equiv  -(W^v_1)^{k-2}(U_0(h)) $.
In this case, $U^v_k$'s reduces to the left-invariant vector fields
$W^v_k \equiv  ({\cal L}_{W^v_1})^{k-1}(U_0)$ which are
$\rho_{\varphi}$-divergence-free.

\item
$U_k, U^h_k$ and $W^v_k$ are Hamiltonian vector fields with
the common Hamiltonian function given by generalized potential
vorticities $h_k$.

\item
The Poisson bracket
algebra of generalized potential vorticities
$h_k$'s is isomorphic to the algebra of vector
fields $W^v_k$ generating the time-dependent particle relabelling
symmetries on $M$.

\item
$U^h_k$'s form the commutative gauge algebra of reparametrization
of trajectories of $v$ on $M$.

\item
This framework for three dimensional flows is
analogous to the natural symplectic structure of two dimensional
one and enables us, as in two dimensions, to represent the generators
of volume preserving diffeomorphisms by Hamiltonian vector fields
and thereby to express the vorticity bracket through the Poisson bracket
of invariant functions on $M$.
\end{itemize}

The geometric character and the algebraic consequences
of this construction and classification of
symmetries distinguish the present work
from the approaches that employ the analysis of defining
equations as the main tool \cite{hen},\cite{sagdeev},\cite{kuzmin}.
Such constructions require that
the algebraic structure of the solution set
consisting of symmetries and invariants be separately treated. Instead,
we gave a recursive construction of symmetries via the Jacobi identity
and, moreover, realized them as the Hamiltonian
automorphisms of the symplectic two-form. This
ensures that the infinitesimal symmetries thus obtained
constitute a Lie algebra
because, although not every infinite set of vector fields form
a Lie algebra, Hamiltonian vector fields do so.

We showed that some properties of two dimensional flows
derived from their symplectic structure can be realized
in three dimensions as well. However, there still remains
the problem of constructing infinitely many helicity type
integrals analogous to the enstropy type Casimirs (\ref{cas})
of two dimensional flows \cite{shep}.
Using the present results, we shall elobarate, in a forthcoming
publication, the kinematical
conservation laws associated to the particle relabelling symmetries.
These invariants are related to the problem
of orbits of coadjoint action of $Diff_{vol}(M)$ and describe
the reduced phase space of the Eulerian dynamical equations
\cite{via},\cite{mw}.
We, therefore, expect the present work to be useful in advencing in 
the general problem of orbits of level sets of momentum mapping
for systems with infinitely many kinematical symmetries.

\newpage

\section*{Acknowledgements}

Jerry Marsden has introduced me, while I was a postdoc with him
and since then through his published works,
to the geometry of fluids and plasmas.
The present work was initiated from a question raised by
the referee of \cite{hg97}.
I thank Yavuz Nutku, Ay\c{s}e Erzan, Itamar Procaccia and Alp Eden
for encouragements and interests.
I acknowledge various inspirational conversations with Kamuran
Sayg{\i}l{\i}, \.Ilhan \.Ikeda and Andrei Ratiu.
I am indepted to B\"{u}lent Karas\"{o}zen for being a patient director
of a library at a distance.

\end{document}